\documentclass[aps,superscriptaddress,onecolumn,pre,longbibliography,floatfix,notitlepage]{revtex4-1}

\usepackage{amsmath,amsthm,amsfonts,amssymb,bm,graphicx,color,times}
\usepackage[colorlinks={true}, citecolor={blue}, filecolor={blue}, linkcolor={blue}, urlcolor={blue}]{hyperref}
\usepackage[caption=false]{subfig}
\usepackage{dsfont}
\usepackage{indentfirst}
\usepackage{soul}
\usepackage{ulem}

\newcommand{\rr}{\textcolor{black}}
\begin{document}

\title{Black-hole powered quantum coherent amplifier}

\author{Avijit Misra}
\email{avijit.misra@weizmann.ac.il}
\affiliation{AMOS and Department of Chemical and Biological Physics,
Weizmann Institute of Science, Rehovot 7610001, Israel}
\author{Pritam Chattopadhyay}
\email{pritam.chattopadhyay@weizmann.ac.il}
\affiliation{AMOS and Department of Chemical and Biological Physics,
Weizmann Institute of Science, Rehovot 7610001, Israel}
\author{Anatoly Svidzinsky}
\email{asvid@physics.tamu.edu}
\affiliation{Texas A\& M University, College Station, Texas 77843, USA}
\author{Marlan O. Scully}
\email{scully@tamu.edu}
\affiliation{Texas A\& M University, College Station, Texas 77843, USA}
\affiliation{Baylor University, Waco, Texas 76798, USA}
\affiliation{Princeton University, Princeton, New Jersey 08544, USA}
\author{Gershon Kurizki}
\email{gershon.kurizki@weizmann.ac.il}
\affiliation{AMOS and Department of Chemical and Biological Physics,
Weizmann Institute of Science, Rehovot 7610001, Israel}
\date{\today}
\begin{abstract}
Atoms falling into a black hole (BH) through a cavity are shown to enable
coherent amplification of light quanta powered by the BH
gravitational vacuum energy. This process can harness the BH energy towards useful purposes, such as propelling a
spaceship trapped by the BH. The process can occur via transient amplification of a signal
field by falling atoms that are partly excited by Hawking radiation reflected by an
orbiting mirror. In the steady-state regime of thermally equilibrated atoms
that weakly couple to the field, this amplifier constitutes a BH-powered
quantum heat engine. The envisaged effects substantiate the thermodynamic approach to BH
acceleration radiation.
\end{abstract}

\maketitle

\begin{figure*}
\centering
\includegraphics[scale=0.46]{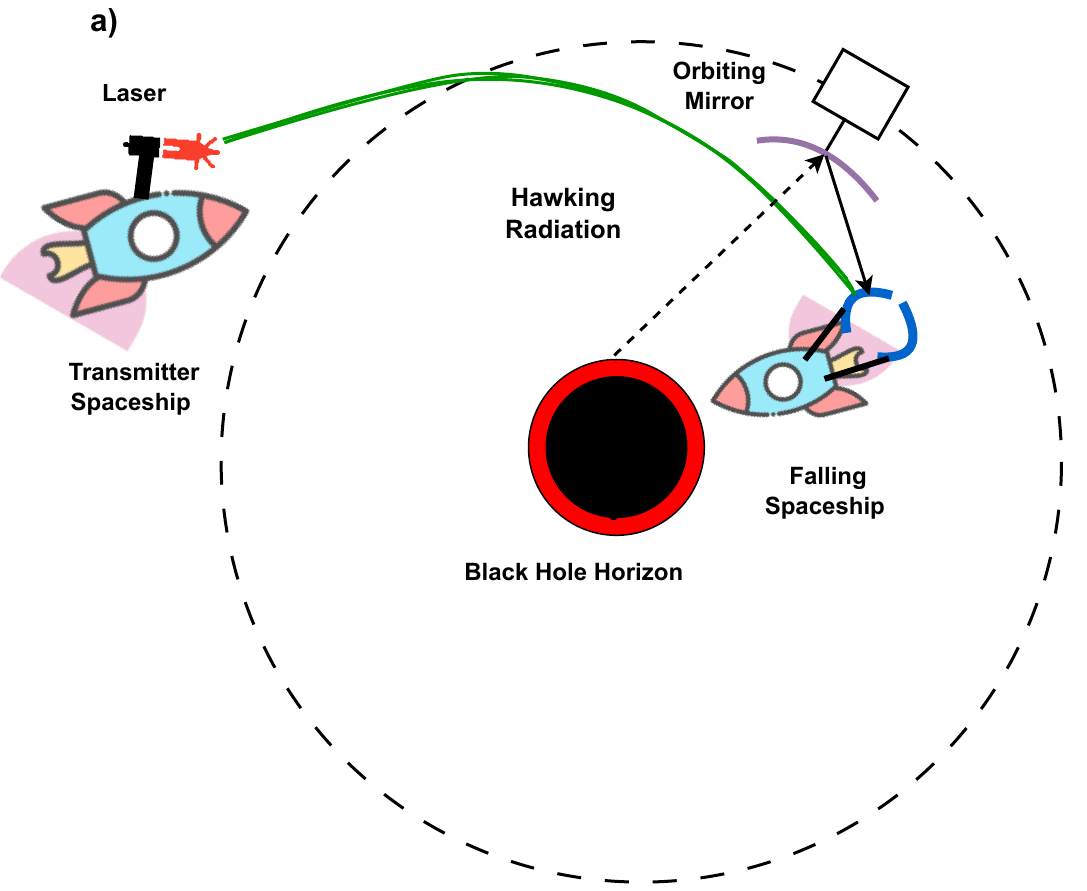}
 \includegraphics[scale=0.65]{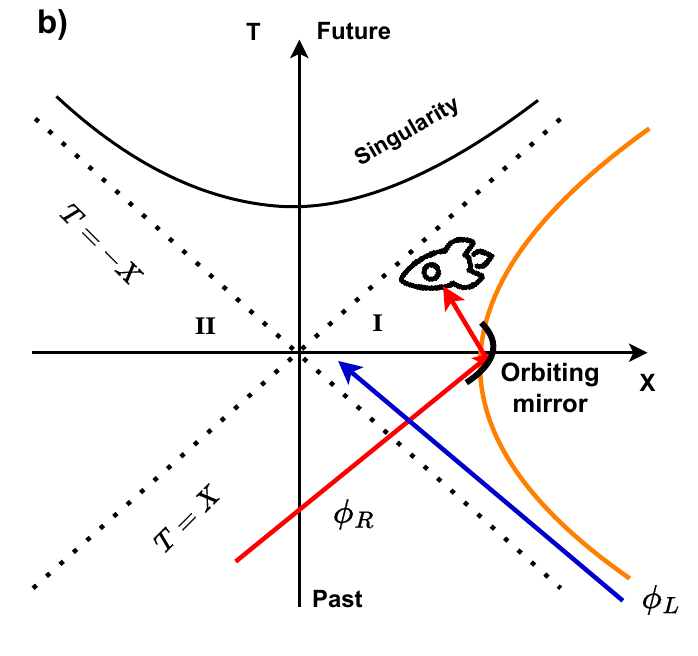}
 \includegraphics[scale=0.6]{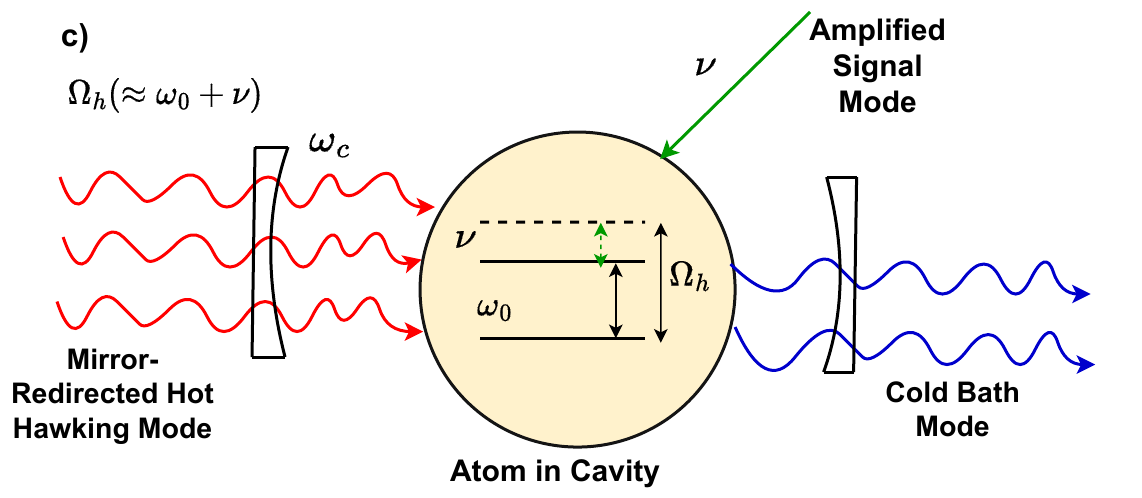}
\caption{a) Coherently amplified energy transfer between spaceships is enabled
by a cavity filled with atoms mounted on a spaceship that is freely falling into a BH provided the BH radiation is redirected by an orbiting mirror. 
 b) Space-time diagram of the relevant scalar modes in the Kruskal coordinates. c) Schematic description of the amplification process in the cavity.}
% by the cavity with atoms in the satellite that is orbiting around the black hole.}
\label{fig:my_label1}
\end{figure*}

\textbf{Introduction:}
Imagine a scene that can play out in a science fiction movie (Fig.~\ref{fig:my_label1}): a spaceship is helplessly falling into a black hole (BH) because its fuel supply is
dwindling and does not suffice for a breakaway maneuver. Luckily, its SOS
message has been received by a faraway spaceship, which is equipped with a
powerful laser that can transfer coherent energy to its distressed sister
ship. Unlike heat, coherent energy transfer is associated with \textit{ergotropy} \cite{allahverdyan2004maximal,pusz1978passive,ErgoPRL, Us_natcom,WOF,Us_pre,francica2017daemonic,2021,singh2021partial,perarnau2015extractable,PhysRevA.104.L030402} that can
perform mechanical work~\cite{book1} to propel the ship. Unfortunately,
coherent energy transfer would have poor efficiency due to diffraction and
BH gravitational lensing over large distances between the ships. Yet a revolutionary technique may still rescue the
ill-fated spaceship: the laser signal can be coherently amplified in a novel
fashion by atoms in free fall through a cavity. Namely, the amplification can only occur
through excitation of the free-falling atoms by BH Hawking radiation redirected by an orbiting mirror.
The envisioned amplification can strongly enhance the coherent power
transfer to the falling spaceship, providing it with enough thrust to free itself
from the grip of the BH.

What is the theoretical basis for this fantastic story? It is the
mind-boggling idea that the Unruh vacuum~\cite{PNAS,PRD1,PRD2,Wald} yields thermal Hawking radiation near the
BH horizon, but cannot directly excite atoms falling into the BH, % if this radiation is redirected towards the BH, say, by a mirror that
%orbits the BH~\cite{PNAS}. This excitation process is unique, since 
as opposed to a bright
star that can directly heat up falling atoms in its vicinity.
By contrast, near a BH the free-falling atoms feel the heat only if the Hawking
radiation is \textit{redirected by a mirror} placed on a stable orbit around the BH (Fig.~\ref{fig:my_label1}). 

Then, counter-intuitively, BH
gravity can act on atoms as a heat bath, although the process is purely
unitary \cite{PNAS,PRD1,PRD2,Wald,sen2022equivalence,mitra2020binary}.

For atoms falling into a BH during their passage through a cavity, a
perturbative (master-equation) approach maps this BH-gravitational problem onto
that of a \textit{quantum heat engine} that acts as a \textit{two-level maser/laser without population inversion}
coupled to two baths at different temperatures \cite{Ghosh2018}. Here the piston of the heat engine is the signal laser field whereas the BH scalar field modes redirected by a mirror replace the hot bath as the energy source and the cold bath as the entropy dump of the
engine.  This uniquely quantum mechanical manifestation of anomalous,
gravitational vacuum effect unequivocally
demonstrates the validity of the thermodynamic approach to acceleration
radiation near a BH. Another intriguing limit is the strong-coupling field-atom regime mediated by the BH vacuum state, a novel manifestation of gravity-induced quantum electrodynamics.

\textbf{Analysis:}
% \begin{figure}[th]
% \centering
% \includegraphics[scale=0.48]{BH_1aa.pdf}
% \caption{Schematic description of the physical process}
% \label{fig:my_label_x}
% \end{figure}
A cloud of two-level atoms (TLA) initially in their ground state, is freely
falling towards the BH through a cavity.
%with resonance frequency $\omega _{c}$. 
The TLA are coupled to the gravitational field of the BH by a quantized
scalar field \cite{PNAS,PRD1,PRD2,Wald} 
\begin{equation}
\hat{\Phi}(\Vec{r},t)=\sum_{i}[\hat{a}_{i}\phi _{i}(\Vec{r},t)+H.c.],
\end{equation}%
where \textit{H.c.} stands for the Hermitian conjugate, index $i$ labels the
field modes, $\Vec{r}=(r,\Theta )$ denotes the radial and angular
coordinates, and $\hat{a}_{i}$ is the $i-$th mode annihilation operator.

The scalar field is coupled with the TLA as depicted in the space-time
diagram (Fig.~\ref{fig:my_label1}b). An atom freely falling into a non-rotating BH while still above
the horizon can (see App.~\ref{AppendixB}) be resonant with the following scalar
field modes (in the Kruskal-Szekeres coordinates) %\cite{PNAS,PRD1,PRD2}  
\begin{equation}
\phi _{1\Omega }(T,X)=e^{-i\Omega \left( T-X\right) },  \label{p1}
\end{equation}%
\begin{equation}
\phi _{2\Omega }(T,X)=\left( T+X\right) ^{-i\Omega }\theta (T+X),  \label{p2}
\end{equation}%
where $\theta $ is the step function and $\Omega >0$. From the perspective
of the free-falling atom the modes (\ref{p1})-(\ref{p2}) harmonically
oscillate as a function of the atom's proper time with positive frequency.
The form of the outgoing mode (\ref{p1}) and the ingoing mode (\ref{p2}) derived here (App. \ref{AppendixB}) is, as shown below,  key to our 
ability to employ  the BH as a source of useful quanta. 
% the modes $%\phi _{1\Omega }$ and $\phi _{2\Omega }$ have positive frequency. 

The free-falling atoms may resonantly interact with the outgoing
plane-wave field $\phi _{1\Omega }$ and with the ingoing Rindler field $\phi
_{2\Omega }$. However, in the Unruh vacuum, which by consensus represents
the state of the evaporating BH field \cite{Wald}, there are no photons in the modes (%
\ref{p1}) and (\ref{p2}). Consequently, \textit{free-falling atoms cannot
become excited in the Unruh vacuum} (see App. \ref{AppendixB}).

Instead, we might consider exciting these atoms by the \textit{outgoing} Rindler photons, which fill the Unruh vacuum and constitute the Hawking radiation \cite{Hawk74,Hawk75}. They thermally populate the modes
\begin{equation}
\phi _{3\Omega }(T,X)=\left( X-T\right) ^{i\Omega }\theta (X-T).  \label{p4}
\end{equation}Yet, it can be shown (App. \ref{AppendixB}) that these outgoing Rindler photons cannot excite free-falling atoms. Is there another way to excite these atoms by BH radiation?

Indeed, there is: we show that free-falling atoms can be excited by redirecting the outgoing Rindler photons (Hawking
radiation) towards the BH via a mirror. The mirror should orbit the BH at
a fixed radius $r=r_{0}$. %which can be far from the horizon. 
To be stable, the mirror orbit should lie at $r > 3r_g$, $r_g$ being the gravitational radius, but otherwise the value of $r$ does not affect the result (see below). 
In the presence of such
a mirror, the mode function satisfying the boundary condition $\phi
(t,r_{0})=0$ at the mirror surface acquires a new, advantageous form 
\begin{equation}
\phi (T,X)=\underbrace{\left( X-T\right) ^{i\Omega }}_{\text{$\phi _{c}$ mode}}-\underbrace{e^{i\Omega \left( r_{0}+\ln (r_{0}-1)\right)
}\left( T+X\right) ^{-i\Omega }}_{\text{$\phi _{h}$ mode}}.  \label{w1}
\end{equation}%
This hitherto \textit{unexplored scalar field mode} has two parts: the outgoing Rindler
photon mode (the first term on the rhs) and a part reflected from the mirror
into the ingoing Rindler mode (the second term on the rhs). This ingoing
Rindler mode acts as a hot bath mode, denoted as $\phi _{h}(r,t)$ with frequency $%
\Omega =\Omega _{h}$, that can excite the free-falling atom. The outgoing Rindler modes act as a cold-bath (vacuum state) mode denoted as $\phi _{c}(r,t)$.

We wish to show that the
redirected Hawking radiation can enable coherent
amplification of a signal mode. 
The complete field-atom interaction Hamiltonian has then the form%
\begin{equation}
H_{int}=\sum_{i}g_{hi}\phi _{hi}\hat{b}^{\dagger }\hat{a}_{hi}|e\rangle
\langle g|+\sum_{j}g_{cj} \phi_{cj} \hat{c}_{cj}|e\rangle \langle g|+H.c.  \label{sc1}
\end{equation}
%  \omega_0 |e\rangle\langle e| + \nu b^\dagger b + \sum_{i} \Omega_{hi} a_{hi}^\dagger a_{hi} + 
% \sum_{j} \Omega_{cj} c_{j}^\dagger c_{j}       
Here $\hat{b}$ stands for the signal-mode annihilation operator, $ \hat{a}%
_{hi}$ is the i-th mode annihilation operator of the hot bath mode $\phi_{hi}$  of
the redirected Hawking radiation, and $ \hat{c}_{j}$ for that of the j-th
cold bath mode $\phi_{cj}$ %with frequency near $\omega _{c}$ 
of the redirected Hawking radiation (Eq.~\eqref{w1}).

The atom-scalar field interaction (first term on the rhs of Eq. (\ref{sc1}))
represents an \textit{anti-resonant Raman process} whereby a scalar-field quantum in
the $i-$th redirected Hawking-radiation mode $\phi _{hi}$ is converted into
a signal photon by the atomic transition between the ground (g) and excited
(e) states, with coupling strength $g_{hi}$. The interaction Hamiltonian of
the atom with the cold bath $\phi_{cj}$ involves %the product of $\hat{c}_{j}$ and 
the same atomic transition operator $|e\rangle \langle g|$ with coupling
strength $g_{cj}$. Our goal is to maximize the energy gain of the signal mode in a non-passive (ergotropy-carrying)
form, capable of delivering work \cite{book1}.

\vspace{2ex} \textbf{Strong TLA-BH coupling:}
Here we assume that while traversing the cavity, the atom is strongly
coupled to one redirected Hawking radiation mode $\phi _{h}$ with a coupling
strength $g_{h}$ that overwhelms the coupling strengths $g_{cj}$ to all cold
\rr{bath modes, i.e., $g_h >> g_{cj}$. This is justified if the frequency $\omega_c$ of the high-Q cavity mode is taken to be off-resonant with the cold bath mode frequencies, $\Omega_{cj}$, i.e., the detunings $|\Omega_{cj}- \omega_c|$ are large, so that we can neglect the couplings with the cold bath modes.} This scenario corresponds to a high-Q cavity with strong coupling of a single Hawking radiation mode to the atom. To render the problem single-mode, we choose the TLA resonant frequency $\omega _{0}$,
the cavity frequency $\omega _{c}$, the signal $\nu $ and the $\Omega _{h}$
frequency of the redirected mode $\phi _{h}$ in (\ref{w1}) such that \rr{$\Omega _{h}
\approx \nu +\omega _{0} \approx \omega_c$}.
Then the interaction Hamiltonian in Eq.~\eqref{sc1} simplifies to

%Without the cold BH we have: 
%The Hamiltonian of the combined system is
\begin{equation}
H_{int}=g_{h}\phi _{h}\hat{b}^{\dagger }\hat{a}_{h}|e\rangle \langle
g|+H.c.
\end{equation}%
%
% \omega |e\rangle\langle e| + \nu b^\dagger b +  \Omega_{hi} a_{hi}^\dagger a_{hi} +
The basis for the combined atom-field energy states can then be %chosen to be \cite{AM} 
\begin{eqnarray}
|1\rangle &=&|g,n_{s},n_{h}\rangle ,  \notag  \label{sc3-2} \\
|2\rangle &=&|e,n_{s}+1,n_{h}-1\rangle , 
\end{eqnarray}%
where $|n_{s}\rangle $ and $|n_{h}\rangle $ are Fock states of the signal
mode and the BH $\phi _{h}$ mode respectively. %This basis-set is assumed to be non-interacting with other
%sub-spaces, which is only true 
At short times, where first-order transitions
between the atom and the field modes predominate, the subspace in
Eq.~(\ref{sc3-2}) is decoupled from other subspaces, whilst keeping the total
number of excitations constant. 

Let us assume that the atom and the signal mode are initially in the ground and Fock
state $|n_s\rangle$ respectively. \rr{Then}, the initial state of the combined system in general is $\rho^i= |g\rangle
\langle g| \otimes |n_s\rangle \langle n_s|\otimes \rho_{T_c}\otimes \rho_{T_h}$%
, where $\rho_{T_c}$ and $\rho_{T_h}$ are the \rr{cold and hot bath} states at
temperature $T_c$ and $T_h$, respectively. In this problem, $T_c = 0$,   \rr{since, as noted above, the outgoing Rindler modes are in a Minkowski vacuum state that cannot impart photons to the free-falling atom and hence act as zero temperature bath. } The initial state is \rr{then} a mixture of the pure states $|g\rangle
|n_{s}\rangle |n_{h}\rangle $ with probabilities %\begin{equation}
$p_{n_{h}}=e^{-\beta _{h}\Omega _{h}n_{h}}/Z_{\beta _{h}}$, %\end{equation}
where $\beta _{h}=\frac{1}{k_{B}T_{H}}$ is the effective BH (Hawking) temperature \cite{Hawk74,Hawk75}.
\rr{A general analysis of amplification in this regime that allows for non-zero cold-bath temperature (App. \ref{AppendixC}) does not offer conceptually new physics.}

The final-states of the atom and the signal mode after their unitary
evolution over time $t$ are then (App. \ref{AppendixC})
\begin{equation}
\rho_{atom}^f= |u|^2|g\rangle \langle g|+ |v|^2 |e\rangle \langle e|,  \notag
\end{equation}
%the final state of the piston is 
\begin{equation}
\rho_{s}^f= |u|^2|n_s\rangle \langle n_s|+ |v|^2 |n_s+1\rangle \langle n_s+1|
\end{equation}
where

\begin{equation}
u = e^{-\frac{1}{2} i \delta t} \Big(\cos \Big(\frac{1}{2} t \sqrt{\delta
^2+4 \text{g}_h^2 \phi _h^2}\Big)+\frac{i \delta \sin \Big(\frac{1}{2} t 
\sqrt{\delta ^2+4 \text{g}_h^2 \phi _h^2}\Big)}{\sqrt{\delta ^2+4 \text{g}%
_h^2 \phi _h^2}}\Big),  \notag
\end{equation}
\begin{equation}
v = -\frac{2 i \text{g}_h \phi _h e^{-\frac{1}{2} i \delta t} \sin \left(%
\frac{1}{2} t \sqrt{\delta ^2+4 \text{g}_h^2 \phi _h^2}\right)}{\sqrt{\delta
^2+4 \text{g}_h^2 \phi _h^2}},
\end{equation}

\begin{equation}
\delta = \omega_0 + \nu - \Omega_h.  \notag
\end{equation}

The \textit{work capacity (ergotropy)} \rr{increase (gain)} following the interaction in the cavity is (App. \ref{AppendixD}, \ref{AppendixC})
\begin{equation}
\mathcal{W}(\rho _{s}^{f})-\mathcal{W}(\rho _{s}^{i})=\nu (|v|^{2}-|u|^{2}),
\end{equation}%
which \rr{requires $|v| > |u|$ and} is maximized for $|v|=1$, $|u|=0$.

For the choice $\delta =0$, $\text{g}_{h} t|\phi _{h}|=(2m+1)\pi /2$, where $%
m $ is an integer, the atom is transferred to the excited state and the
signal adds a photon to its mode, $\rho _{s}^{f}=|n_{s}+1\rangle \langle n_{s}+1|$. The highest amplification per atom is achieved for $n_s = 1$.  The
efficiency of work extraction by the signal from the BH is then 
\begin{equation}\label{SSDeq1}
\eta = \frac{\nu }{\omega _{0}+\nu }.
\end{equation}%
This efficiency can closely approach the Scovil-Schulz-Dubois (SSD) bound of quantum heat engine/amplifiers~%
\cite{SSD} $\nu /(\omega _{0}+\nu )$. In turn, the SSD efficiency $\eta_{\rm SSD}$ can approach the Carnot efficiency $\eta_C$ if $\frac{T_h}{T_c} \gtrsim \frac{\Omega_h}{\omega_c}$. However, as $T_c \rightarrow 0$,  the atom resonant frequency must approach zero in order to attain the Carnot efficiency, which is unfeasible.

The maximal average \rr{output} power in this regime \rr{is given by the time derivative of the ergotropy (work extraction) increase}
\begin{equation}
    \mathcal{\dot{W}}= \frac{ 2 g_h |\phi_h|\nu }{(2m+1)\pi},
\end{equation}
where the maximal power corresponds to $m=0$.

Spectacular {\it power boost} can be obtained in the Dicke regime of $N$ atoms that are collectively coupled to the hot bath mode. Following~\cite{niedenzu2018cooperative}, we can have
\begin{equation}
    \mathcal{\dot{W}} \rightarrow N \mathcal{\dot{W}}.
\end{equation}
 %(see Eq.~\eqref{eq12345}).

\vspace{2ex} \textbf{Weak TLA-BH coupling:} 

Let us now consider the opposite limiting regime of a cavity with insufficiently high Q and \rr{resonance frequency $\omega_c \approx \omega_0$.}  \rr{Also the detuning $|\omega_c-\Omega_{cj}|$ of the cavity frequency with the cold bath modes is small, so that the atom equilibrates with the cold bath modes $\phi_{cj}$ at $T_C=0$ and reaches its ground state $|g\rangle$, following the interaction (\ref{sc1}) which is now dominated by
\begin{equation}
H_{\mbox{int}}\approx\sum_{j}(g_{cj} \phi_{cj} \hat{c}_{cj}|e\rangle \langle g|+H.c.)  \label{sc-weak}
\end{equation}}
%\rr{such that the dominating contribution of the interaction  Hamiltonian in Eq. (\ref{sc1}) comes from the cold bath mode $\phi_c$ over the redirected Hawking radiation mode $\phi_h$, which reads as
%\begin{equation}
%H_{int}=g_{c} \phi_{c} \hat{c}_{c}|e\rangle \langle g|+H.c.  \label{sc-weak}
%\end{equation}}
\rr{Though the atom reaches its steady state by the interaction with the cold bath modes, the atom together with the signal mode are energized by the redirected Hawking radiations via coupling to the $\phi_{hi}$ modes (where $\phi_{hi}\approx \omega_0+\nu$).}
%\rr{as 
%\begin{equation}\label{abc11}
 % \frac{\rho_{ee}}{\rho_{gg}} \approx \frac{\bar{n}_c}{\bar{n}_c+1}= %\exp{\left[-\frac{\hbar \omega_c}{k_B T_c}\right]},   
%\end{equation}
%$\rho_{ee}$ and $\rho_{gg}$ being respectively the excited and ground state populations of the atom. Here $\bar{n}_c=\frac{1}{\exp[-\hbar \omega_c/k_B T_c]-1}$ is the mean number of photons in the cold bath mode at temperature $T_c$.} \bb{In this case $T_C=0$ and the atom relaxes to the ground state steadily.
%}
 The process is analogous to our continuously
operating heat-engine maser based on a TLA \cite{Ghosh2018}. \rr{This treatment yields the effective  Raman Hamiltonian for the atom-scalar field coupling that reads in the interaction picture.}
%Here, the atom
%together with the signal at frequency $\nu $ are coupled to a hot field mode near resonantly, but the coupling strength $g_{h}$ is assumed to be weaker than the coupling to the cold modes $g_{c}$. 

% The atom-scalar field interaction
% obeys the Raman Hamiltonian that in the interaction picture reads (
% \cite{Ghosh2018}) 
\begin{equation} \label{wcoup1}
H_{\mbox{Raman}}(t)=g_{h}\sum_{i}\left( \phi _{hi}\hat{a}_{hi}\hat{b}^{\dagger
}|e\rangle \langle g|e^{-i[\Omega _{hi}-(\nu +\omega _{0})]t}+H.c.\right) .
\end{equation}
Under this interaction, we get a master equation for the state of the
hot scalar field (App. \ref{AppendixA}). By tracing out the atom, which has reached a steady
state under the influence of the cold bath, we then find the time
evolution of the signal mode \rr{(\cite{Ghosh2018})}.

\begin{figure}[ht]
\centering
\includegraphics[scale=0.25]{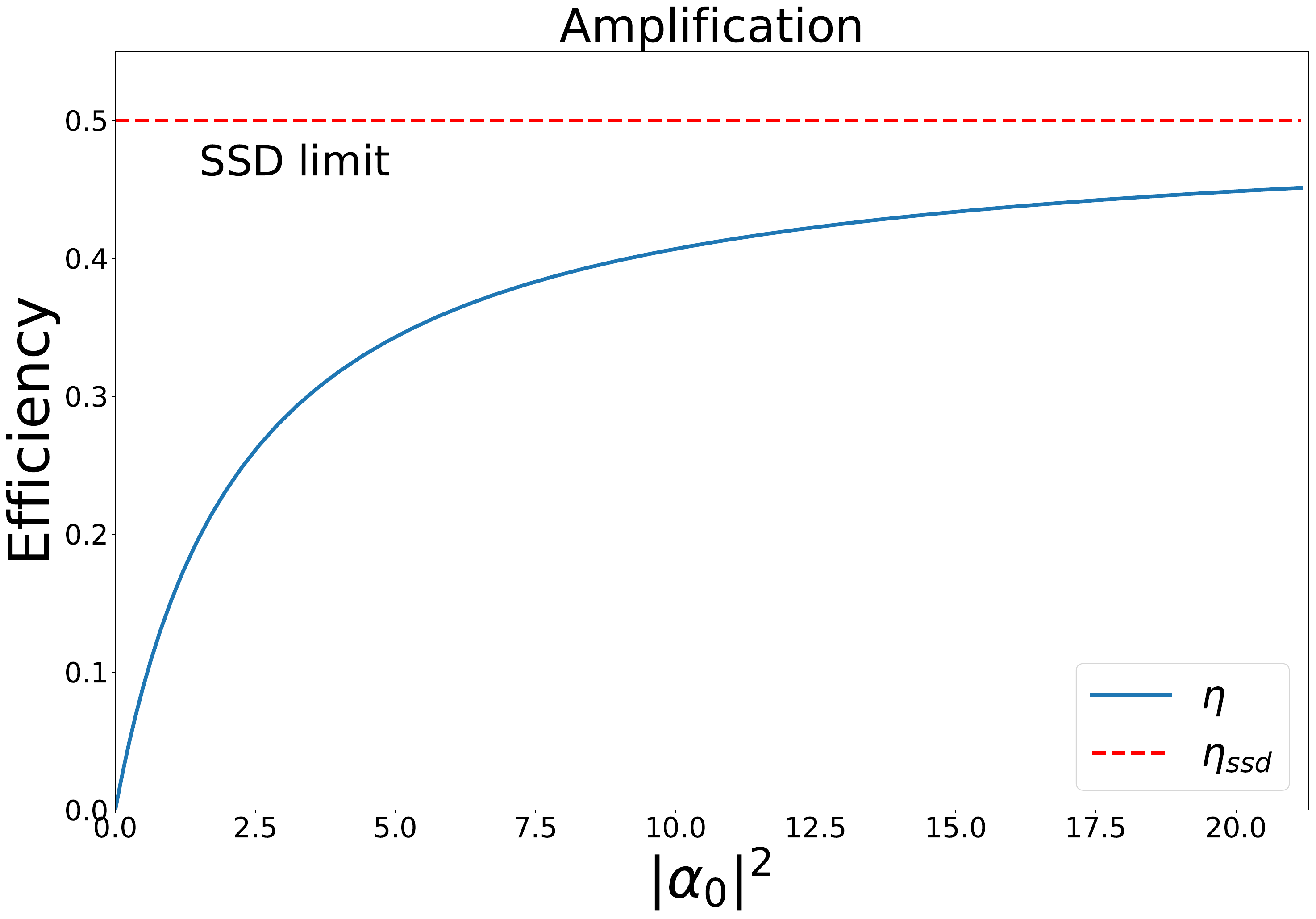}
\caption{Efficiency $\protect\eta$ of the amplification process in terms of
the initial mean squared amplitude of the signal $|\protect\alpha_0|^2$. For
large $|\protect\alpha_0|^2$ the Scovil-Schulz-Dubois (SSD) limit is
attained. }
\label{fig:my_label}
\end{figure}

\begin{figure}[th]
\centering
\includegraphics[scale=0.25]{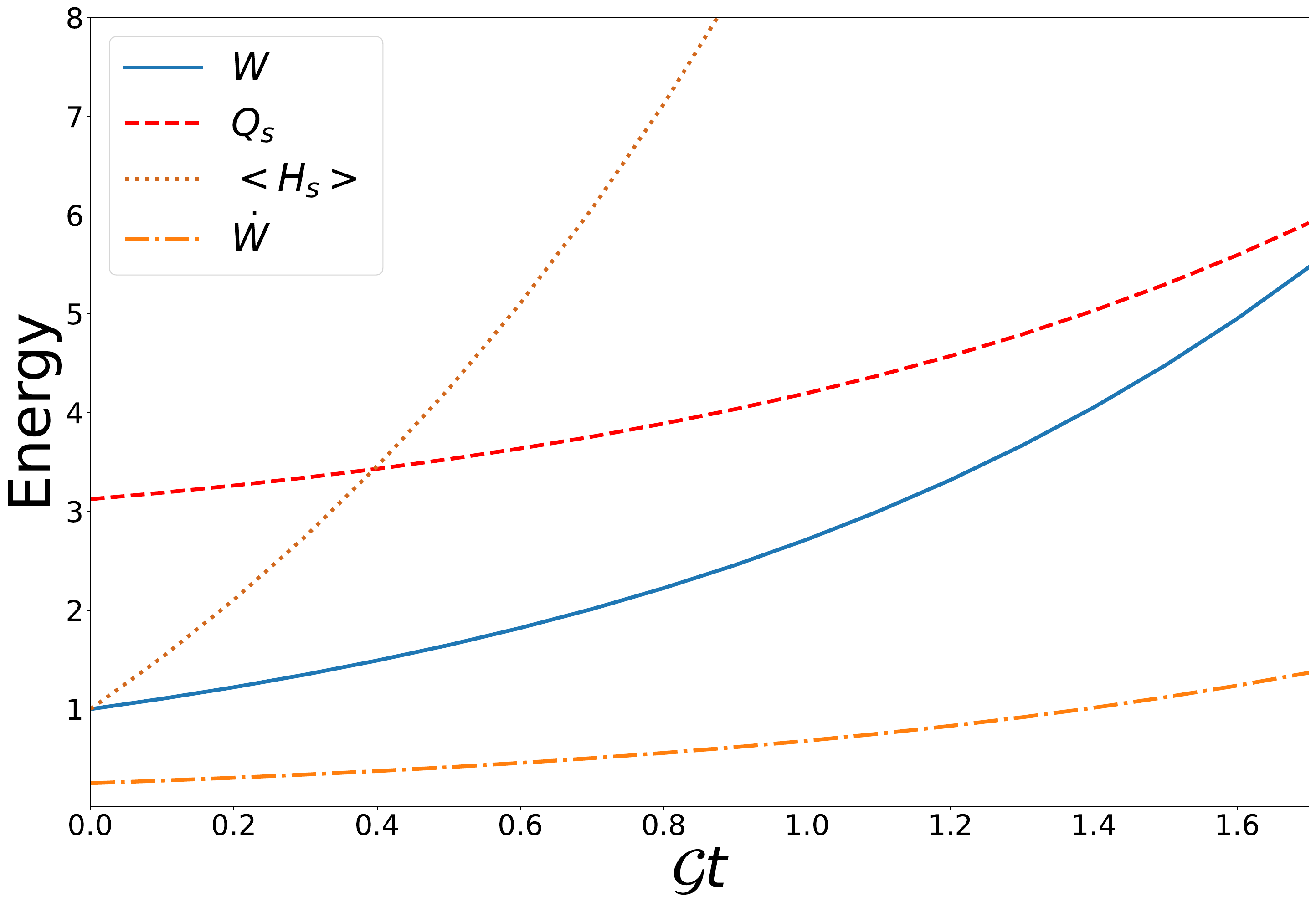}
\caption{ Work capacity, ergotropy, thermal energy, and the mean energy of
the signal field induced by the BH as a function of the amplifier gain $%
\mathcal{G}$ (in inverse time units).}
\label{fig:my_label123}
\end{figure}

The ergotropy (work capacity) of the signal state, \rr{initially in a coherent state $|\alpha_0\rangle$}, undergoes coherent amplification as (App. \ref{AppendixA}) %grows with time as
\begin{equation}
    \mathcal{W}= \nu |\alpha_0|^2 e^{\mathcal{G}t},
\end{equation}
 where $|\alpha _{0}|$ is the mean initial signal amplitude and $\mathcal{G}$ is the gain \rr{(\cite{Ghosh2018})}. The power of the gained work is therefore given by
 \begin{equation}
    \dot{ \mathcal{W}}= \mathcal{G} \nu |\alpha_0|^2 e^{\mathcal{G}t}.
 \end{equation}

As in the strong-coupling regime, $N$-fold collective (Dicke) power boost~\cite{niedenzu2018cooperative} is attainable by $N$ atoms. 

The efficiency can be computed as the ratio of power
generated by the signal %$\dot{W}$ 
%$= \hbar \nu \mathcal{G} |\alpha_0|^2 e^{\mathcal{G}t}$ 
to the heat flux from the BH, $\dot{Q}_{h}$, \rr{which is the rate of energy change of the working system induced by the heat bath effect~\cite{book1,breuer2002theory}. 
%%Mathematically, 
%\begin{equation}
 %   \dot{Q}_{B} = \text{Tr} (\dot{\rho}_B H_{\mbox{WS}}),
%\end{equation}
%where $B$ denotes the bath we are interested in, $\dot{\rho}_B$ is the time evolution of the working system induced by the specific bath $B$ and $H_{\mbox{WS}}$ being the Hamiltonian of the working system. 
In our scheme,
$\dot{Q}_{h} = \text{Tr} (\dot{\rho}_h H_{\mbox{TLA+Signal}})$, where $H_{\mbox{TLA+Signal}}= \omega_0 |e\rangle\langle e|+\nu b^\dagger b$ is the free (combined) Hamiltonian of the TLA and the signal mode.}
%$\dot{Q}_h= \hbar \Omega_h (\mathcal{G} \langle b^\dagger b \rangle + \mathcal{D})$. 
\rr{Given that $T_c =0$,} the efficiency \rr{then} evaluates \rr{to} (App. \ref{AppendixA}), 
\begin{eqnarray}
\eta =\frac{\mathcal{\dot{W}}}{\dot{Q}_{h}} =\frac{\nu }{\Omega _{h}} \frac{|\alpha _{0}|^{2}}{|\alpha _{0}|^{2}+1}.
\end{eqnarray}%
% &=&\frac{\nu }{\Omega _{h}}\frac{|\alpha _{0}|^{2}}{|\alpha |^{2}+\frac{n_{h}(n_{c}+1)}{n_{h}-n_{c}}},
% where $|\alpha _{0}|$ is the mean initial signal amplitude. 
\rr{This efficiency} approaches the Scovil-Schulz-Dubois (SSD) bound $\nu /(\omega _{0}+\nu )$
as $|\alpha _{0}|>>1$ (Fig.\ref{fig:my_label}). In Fig. \ref{fig:my_label123} we show that the division of
the gained signal energy between \rr{increased} ergotropy and \rr{generated} heat tends in favor of ergotropy (coherent work production) as the gain increases.

\vspace{2ex} \textbf{Conclusions:}
We have put forth the possibility of black hole (BH) gravity to act as the
energizing source of coherent light amplification. The amplification is
mediated by the Hawking radiation of the BH in the presence of an orbiting
mirror that transforms outgoing Hawking radiation into ingoing Rindler
quanta. It can be viewed as a BH-fueled heat engine that converts Hawking
radiation into work in a coherent signal mode. 

\rr{We stress that Hawking radiation produced by the black hole (BH) plays a key role in the proposed effect, which can be clarified as follows. Assume that we block the Hawking radiation emanating from the BH event-horizon before it hits the mirror. Then both the outgoing and the ingoing fields are in a Rindler vacuum state, unchanged by the orbiting mirror. In this case the mirror does not generate photons, namely, there is no dynamical Casimir effect~\cite{moore1970quantum}, which would only occur by a mirror accelerated in the Minkowski vacuum, but not in the Rindler vacuum~\cite{rindler1966kruskal}. The free-falling atom can then only be excited upon allowing the Hawking radiation to be redirected back to the BH by the mirror, as we have shown.}

The main energy source in our model is Hawking radiation, and \textit{not the
kinetic or potential energy of the atoms}. In principle, one can also use the
kinetic energy of ground-state atoms passing through the cavity to amplify light \cite{Anatoly}.
Our results corroborate the view \cite{PNAS,PRD1,PRD2,Wald} that, despite the unitarity of such
processes, a BH can act as a heat source on falling matter (cf. \cite{NIHM}).

\rr{The amplifier power is proportional to the flux of Hawking radiation produced by the BH, which scales as $1/M^2$~\cite{bekenstein1980black}, whereas the Hawking (hot-bath) temperature scales as $1/M$. The efficiency is bounded by the Scovil-Schulz-Dubois limit~\cite{SSD} but falls short of the Carnot bound as explained after Eq.~\eqref{SSDeq1}.}

\rr{The efficiency is expected to decrease with the solid angle which the BH spans at the mirror as the orbital distance of the mirror increases, just as in the case of a solar cell whose efficiency depends on the solid angle subtended by the sun~\cite{svidzinsky2021quantum}. However, the decrease of the solid angle as one recedes from the source can be compensated by focusing: As the intensity is increased by focusing, so does the effective solid angle, since the energy flow density per solid angle remains unchanged. Hence, by tightening the focusing one can make the solid angle of the source at the device equal to the solid angle of the radiation emitted by the device, so that the
efficiency becomes independent of the distance.}

Concepts of quantum information theory and optics have been gaining prominence in the context of quantum effects of gravity \cite{QG2,QG1,QG3,membrere2023tripartite}. We here venture in yet another direction, demonstrating that such effects may find practical use, such as propelling a spaceship by atoms falling into a BH. These results open a new avenue that bridges quantum optics, quantum thermodynamics and BH gravity.

\vspace{3ex} \textbf{Acknowledgements: } GK and MOS acknowledge the support
of NSF-BSF. GK acknowledges the support of PACE-IN (QUANTERA), PATHOS (EU
FET OPEN) and DFG (FOR 2724). MOS acknowledges the support of the Air Force
Office of Scientific Research (Grant No. FA9550-20-1-0366 DEF), the Robert
A. Welch Foundation (Grant No. A-1261), and the National Science Foundation
(Grant No. PHY 2013771). 

\vspace{3ex}

\textbf{Author contributions:} GK conceived the initial idea, and then all authors conceptualized and designed the project. AM, PC and AS did the analytical study. PC did the figures and plots. GK and MOS supervised the project. All authors were involved in the analysis and interpretation of the results. GK, AM and AS wrote the manuscript with input from all authors.

\vspace{3ex}

\textbf{Competing interests:} The authors declare no competing interests.

\vspace{3ex}

\textbf{Data availability:} Data sharing is not applicable to this article as no datasets were generated or analyzed during the current study.

\newpage

 \appendix 
%\section{\label{AppendixA} Strong coupling details}

\section{\label{AppendixB} Mode functions of photons resonant with free-falling atoms}

Here we consider a two-level atom with transition frequency $\omega $ freely falling into
a nonrotating BH of mass $M$ along a radial trajectory from infinity with
zero initial velocity. We choose the gravitational radius $r_{g}=2GM/c^{2}$
as a unit of distance and $r_{g}/c$ as a unit of time and introduce the
dimensionless distance, time, and frequency as%
\begin{equation*}
r\rightarrow r_{g}r,\quad t\rightarrow (r_{g}/c)t,\quad \omega \rightarrow
(c/r_{g})\omega.
\end{equation*}

In dimensionless Schwarzschild coordinates the atom trajectory is described
by the equations 
\begin{equation}
\frac{dr}{d\tau }=-\frac{1}{\sqrt{r}},\quad \frac{dt}{d\tau }=\frac{r}{r-1},
\label{u1}
\end{equation}%
where $t$ is the dimensionless time in Schwarzschild coordinates and $\tau $
is the dimensionless proper time for the atom. Integration of equations (\ref%
{u1}) yields%
\begin{equation}
\tau =-\frac{2}{3}r^{3/2}+const,  \label{m3}
\end{equation}%
\begin{equation}
t=-\frac{2}{3}r^{3/2}-2\sqrt{r}-\ln \left( \frac{\sqrt{r}-1}{\sqrt{r}+1}%
\right) +const.  \label{m4}
\end{equation}%
For a scalar photon in the Regge-Wheeler coordinate%
\begin{equation}
r_{\ast }=r+\ln (r-1)  \label{m5}
\end{equation}%
the field propagation equation reads%
\begin{equation}
\left[ \frac{\partial ^{2}}{\partial t^{2}}-\frac{\partial ^{2}}{\partial
r_{\ast }^{2}}+\left( 1-\frac{1}{r}\right) \left( \frac{1}{r^{3}}-\frac{%
\Delta }{r^{2}}\right) \right] \psi =0,  \label{m1}
\end{equation}%
where $\Delta $ is the angular part of the Laplacian. 

We are interested in
solutions of this equation outside of the event horizon, that is for $r>1$.
If the dimensionless photon frequency $\nu \gg 1$, then the first
two terms in Eq. (\ref{m1}) dominate and one can approximately write 
\begin{equation}
\left( \frac{\partial ^{2}}{\partial t^{2}}-\frac{\partial ^{2}}{\partial
r_{\ast }^{2}}\right) \psi =0.
\end{equation}%
The general solution of this equation reads%
\begin{equation}
\psi =F\left( t\pm r_{\ast }\right) =F\left( t\pm r\pm \ln (r-1)\right) ,
\end{equation}%
where $F$ is an arbitrary function. 

We consider a trajectory of the atom near the event horizon and
choose the origin of $\tau $ such that $\tau =0$ when the atom crosses the horizon.
In the vicinity of the horizon, we obtain for the atom's trajectory%
\begin{equation}
t\approx -\ln (-\tau )+\frac{5}{4}\tau +const,
\end{equation}%
\begin{equation}
r\approx 1-\tau -\frac{1}{4}\tau ^{2},
\end{equation}%
and, therefore, along the atom's trajectory 
\begin{equation}
t-r-\ln (r-1)\approx -2\ln (-\tau )+const,  \label{x1}
\end{equation}%
\begin{equation}
t+r+\ln (r-1)\approx \frac{1}{2}\tau +const.  \label{x2}
\end{equation}%
Eqs. (\ref{x1}) and (\ref{x2}) yield the following mode functions of
the field which harmonically oscillates as a function of $\tau $ along the atom's
trajectory%
\begin{equation}
\psi _{1\nu }(t,r)=e^{i\nu e^{-\frac{1}{2}\left( t-r-\ln (r-1)\right)
}}\approx e^{-i\nu \tau },  \label{x3}
\end{equation}%
\begin{equation}
\psi _{2\nu }(t,r)=e^{-2i\nu \left( t+r+\ln (r-1)\right) }\approx e^{-i\nu
\tau }.  \label{x4}
\end{equation}%

It is insightful to write the mode functions (\ref{x3}) and (\ref{x4}) in
the Kruskal-Szekeres coordinates $T$ and $X$ that are defined in terms of
the Schwarzschild coordinates $t$ and $r$ as%
\begin{equation}
T=\sqrt{r-1}e^{\frac{r}{2}}\sinh \left( \frac{t}{2}\right) ,  \label{k1}
\end{equation}%
\begin{equation}
X=\sqrt{r-1}e^{\frac{r}{2}}\cosh \left( \frac{t}{2}\right) ,  \label{k2}
\end{equation}%
for $r>1$, and%
\begin{equation}
T=\sqrt{1-r}e^{\frac{r}{2}}\cosh \left( \frac{t}{2}\right) ,  \label{k3}
\end{equation}%
\begin{equation}
X=\sqrt{1-r}e^{\frac{r}{2}}\sinh \left( \frac{t}{2}\right) ,  \label{k4}
\end{equation}%
for $0<r<1$. In these coordinates, we obtain for $r>1$%
\begin{equation}
e^{-\frac{1}{2}\left( t-r-\ln (r-1)\right) }=X-T,
\end{equation}%
\begin{equation}
T+X=e^{\frac{1}{2}\left( t+r+\ln (r-1)\right) },
\end{equation}%
and, therefore,%
\begin{equation}
\psi _{1\nu }(T,X)=e^{-i\nu \left( T-X\right) },
\end{equation}%
\begin{equation}
\psi _{2\nu }(T,X)=\left( T+X\right) ^{-4i\nu }.
\end{equation}

\section{\label{AppendixD}Ergotropy of the signal mode}
\rr{ To estimate the work gain of the signal mode we resort to ergotropy. The ergotropy of a quantum state $\rho$ with a Hamiltonian $H$ is defined as the maximal amount of average energy that can be extracted by means of a unitary transformation \cite{allahverdyan2004maximal}. It reads
\begin{equation}
    \mathcal{W}(\rho, H)= \mbox{Tr}(\rho H)- \underset{U}{min}\mbox{Tr}(U\rho U^\dagger H).
\end{equation}
The part of the energy that cannot be extracted by a unitary transformations is the passive energy of the state. %and the corresponding state is called a passive state. This passive state of a quantum state is unique with respect to the given Hamiltonian $H$ provided it is non-degenerate. 
For a diagonal density matrix $\rho=\sum p_i |E_i\rangle\langle E_i|$, the state is passive with respect to a Hamiltonian $H=\sum E_i|E_i\rangle\langle E_i|$ only when 
\begin{equation}
    p_i\geq p_j,~~\mbox{if}~~E_j>E_i
\end{equation}
Thus, the total mean energy of a quantum state can be considered as the sum of the ergotropy and passive energy %as given by
\begin{equation}
    E=\mbox{Tr}(\rho H)= E_{\mbox{pas}}+ \mathcal{W}.
\end{equation}
}

\rr{Initially the signal is at $|n_s\rangle\langle n_s|$. By a unitary transformation $|n_s\rangle\rightarrow |0\rangle$, $\nu n_s$ amount of energy of the state can be extracted. Thereby,
the initial ergotropy of the piston mode is given by
\begin{equation}
   \mathcal{W} [\rho_{s}^i]= \nu n_s.
\end{equation}
From the final state of the signal mode (Eq. \ref{fin-sig}) maximal work work extraction is achieved by a unitary transformation: $\{|n_s+1\rangle\rightarrow |0\rangle$; $|n_s\rangle\rightarrow |1\rangle\}$ (when $|v|^2>|u|^2$).
Therefore, the final ergotropy of the signal mode is 
\begin{equation}
   \mathcal{W} [\rho_{s}^f]= \nu [n_s+ (|v|^2-|u|^2)].
\end{equation}
The ergotropy gain or the work gain is 
\begin{equation}
   \mathcal{W}_{\text gain}= \nu (|v|^2-|u|^2),
\end{equation}
which is maximized when $|v|^2=1$.}

\section{\label{AppendixC} Strong-coupling amplifier regime}
The initial state of the combined system is 
\begin{equation}
    \rho^i= |g\rangle \langle g| \otimes |n_s\rangle \langle n_s|\otimes  \rho_{T_h},
\end{equation}
which is a mixture of the pure states $ |g\rangle  |n_s\rangle |n_h\rangle$
with thermal occupation probability of the hot bath mode
\begin{equation}
    p_{ n_h}= e^{-(\beta_h \Omega_h n_h)}/Z_{\beta_h}.
\end{equation}
Each such pure state can be written in the basis in Eq. (\ref{sc3-2}) as 
\begin{equation}
\label{ini-vec}
    |\psi\rangle^i=\left( \begin{array}{c}
         1  \\
         0
    \end{array}\right),
    \end{equation}
    
which under the unitary evolution maps to

   \begin{equation}
  \label{fi-vec}
    |\psi\rangle^f=\left( \begin{array}{c}
    \begin{array}{c}
         e^{-\frac{1}{2} i \delta  t} \Big(\cos \Big(\frac{1}{2} t \sqrt{\delta ^2+4 \text{g}_h^2 \phi _h^2}\Big)\\+\frac{i \delta  \sin \Big(\frac{1}{2} t \sqrt{\delta ^2+4 \text{g}_h^2 \phi _h^2}\Big)}{\sqrt{\delta ^2+4 \text{g}_h^2 \phi _h^2}}\Big)
\end{array} \vspace{3ex}
  
   \& -\frac{2 i \text{g}_h \phi _h e^{-\frac{1}{2} i \delta  t} \sin \left(\frac{1}{2} t \sqrt{\delta ^2+4 \text{g}_h^2 \phi _h^2}\right)}{\sqrt{\delta ^2+4 \text{g}_h^2 \phi _h^2}} 
    \end{array}\right)  = \left( \begin{array}{c}  u \\
         v \end{array}\right).
\end{equation} 
The final state of the atom after time $t$ is then
\begin{equation}
    \rho_{atom}^f=  |u|^2|g\rangle \langle g|+  (|v|^2) |e\rangle \langle e|,
\end{equation}
 and the final state of the piston is 
\begin{equation}\label{fin-sig}
    \rho_{p}^f= |u|^2|n_s\rangle \langle n_s|+  |v|^2 |n_s+1\rangle \langle n_s+1|.
\end{equation}
Here we have taken the sum over all pure state in Eq. (\ref{ini-vec}) with the thermal probability $p_{ n_h}$ in the hot bath mode.

\rr{\textbf{Effect of the cold bath:-} The basis states for the signal-atom-BH system combined along with the cold bath mode is  
\begin{eqnarray} \nonumber \label{sc3}
    |1\rangle & = & |g, n_s,n_c,n_h\rangle, \\ \nonumber
    |2\rangle & = & |e, n_s,n_c-1,n_h\rangle, \\
    |3\rangle & = & |e, n_s+1,n_c,n_h-1\rangle.
\end{eqnarray}
Then the initial state of the combined system+bath is a mixture of the pure states $ |g\rangle\langle g|  |n_s\rangle\langle n_s| |n_c\rangle \langle n_c| |n_h\rangle \langle n_h|$ with probability
\begin{equation}
   p_{n_c n_h}= e^{-(\beta_c \Omega_c n_c+\beta_h \Omega_h n_h)}/Z_{\beta_c}.Z_{\beta_h}.
\end{equation}}
\rr{In the basis in Eq. (\ref{sc3}), the initial state vector for the ${g,n_s,n_c,n_c}$  subspace is
\begin{equation}
\label{ini-vec}
   |\psi\rangle_{g,n_s,n_c,n_c}^i=\left( \begin{array}{c}
        1  \\
        0\\
        0 
   \end{array}\right).
   \end{equation}
    Under the unitary evolution in Eq. (6) in the main text, it evolves to
\begin{equation}
 \label{fi-vec}
   |\psi\rangle^f=\left( \begin{array}{c}
   \begin{array}{c}
        e^{-\frac{1}{2} i \delta  t} \Big(\cos \Big(\frac{1}{2} t \sqrt{4 \text{g}_c^2 \phi _c^2+\delta ^2+4 \text{g}_h^2 \phi _h^2}\Big)\\+\frac{i \delta  \sin \Big(\frac{1}{2} t \sqrt{4
  \text{g}_c^2 \phi _c^2+\delta ^2+4 \text{g}_h^2 \phi _h^2}\Big)}{\sqrt{4 \text{g}_c^2 \phi _c^2+\delta ^2+4 \text{g}_h^2 \phi _h^2}}\Big) \vspace{3ex}
\end{array}\\ 
 -\frac{2 i \text{g}_c \phi _c
  e^{-\frac{1}{2} i \delta  t} \sin \left(\frac{1}{2} t \sqrt{4 \text{g}_c^2 \phi _c^2+\delta ^2+4 \text{g}_h^2 \phi _h^2}\right)}{\sqrt{4 \text{g}_c^2 \phi _c^2+\delta ^2+4 \text{g}_h^2
  \phi _h^2}} \vspace{3ex} \\ 
  -\frac{2 i \text{g}_h \phi _h e^{-\frac{1}{2} i \delta  t} \sin \left(\frac{1}{2} t \sqrt{4 \text{g}_c^2 \phi _c^2+\delta ^2+4 \text{g}_h^2 \phi _h^2}\right)}{\sqrt{4
  \text{g}_c^2 \phi _c^2+\delta ^2+4 \text{g}_h^2 \phi _h^2}} 
   \end{array}\right)  = \left( \begin{array}{c}  a  \\
        b\\
        c  \end{array}\right).
\end{equation} 
The final state of the atom after time $t$ is then
\begin{equation}
   \rho_{atom}^f=  |a|^2|g\rangle \langle g|+  (|b|^2+|c|^2) |e\rangle \langle e|
\end{equation}
 and the final state of the signal is 
\begin{equation}
   \rho_{s}^f=  (|a|^2+|b|^2)|n_s\rangle \langle n_s|+  |c|^2 |n_s+1\rangle \langle n_s+1|
\end{equation}
We can have the atom in the excited state by requiring $|a|^2=0$ which holds for resonant transition $\delta=0$, and 
\begin{equation}
  \Big(\frac{1}{2} t \sqrt{4 \text{g}_c^2 \phi _c^2+\delta ^2+4 \text{g}_h^2 \phi _h^2}\Big)= (2n+1) \pi/2,
\end{equation}
where $n \in \mathds{N}$. 
For the same choice, the state of the signal is 
\begin{equation}
   \rho_{s}^f= |b|^2 |n_s\rangle \langle n_s|+  |c|^2 |n_s+1\rangle \langle n_s+1|.
\end{equation}
To have the signal in a population inverted state we need $|c|^2>|b|^2$, which is only possible when   
\begin{equation}
  |\text{g}_h \phi _h| > |\text{g}_c \phi _c|
\end{equation}
The ergotropy or the work gain (when $a=0$ and Eq. B19 holds ) is then given by 
\begin{equation}
   \mathcal{W}_{\text gain}= \nu (|c|^2-|b|^2). 
\end{equation}
The efficiency in this case is given by 
\begin{eqnarray}
  \eta &=&  \frac{\nu}{\omega_0+ \nu}[1-\frac{|b|^2}{|c|^2}] \nonumber \\ 
  &=&  \frac{\nu}{\omega_0+ \nu}[1-\frac{(g_c \phi_c)^2}{(g_h \phi_h)^2}].
\end{eqnarray}
In the limit $|g_{c} \phi_c| \rightarrow 0$, we retrieve the results of the two modes+TLA case in the main text.}

\section{\label{AppendixA} Weak-coupling amplifier regime}
\rr{In this regime, the atom reaches a steady state  under the action of the cold bath as
\begin{equation}\label{abc11}
  \frac{\rho_{ee}}{\rho_{gg}} \approx \frac{\bar{n}_c}{\bar{n}_c+1}\approx \exp{\left[-\frac{\hbar \omega_0}{k_B T_c}\right]},   
\end{equation}
$\rho_{ee}$ and $\rho_{gg}$ being respectively the excited and ground state populations of the atom. Here $\bar{n}_c=\frac{1}{\exp[\hbar \Omega_c/k_B T_c]-1}$ is the mean number of photons in the cold bath mode at temperature $T_c$.} 
Then the master equation (ME) for the combined signal-atom state associated with the hot bath mode is~\cite{book1,Ghosh2018}

\begin{eqnarray}
\dot{\rho}_h&=& g_h^2 |I_{h,gi}|^2 (\bar{n}_h+1)([S\rho_h,
S^\dagger]+[S,\rho_h S^\dagger])  \notag \\
&+& g_h^2 |I_{h,ei}|^2 \bar{n}_h([S^\dagger\rho_h, S]+[S^\dagger,\rho_h S]),
\end{eqnarray}
where $S=b |g\rangle \langle e|$, $\Bar{n_h}$ is the mean quanta number in
the thermal state associated with the Hawking radiation, and 
\begin{eqnarray}
|I_{h,gi}|^2= \int_{t_i} ^{t_f} dt^\prime e^{-i \delta_{ci} t^\prime}
\phi_h^{\star} (t^\prime) \int_{t_i} ^{t_f} dt^{\prime \prime} e^{i
\delta_{ci} t^{\prime \prime}} \phi_h (t^{ \prime \prime}) ,  \notag \\
|I_{h,ei}|^2= \int_{t_i} ^{t_f} dt^\prime e^{i \delta_{ci} t^\prime} \phi_h
(t^{ \prime}) \int_{t_i} ^{t_f} dt^{\prime \prime} e^{-i \delta_{ci}
t^{\prime \prime}} \phi_h^{\star} (t^{ \prime \prime}),
\end{eqnarray}
where $\delta_{ci} =(\Omega_{ci}- \omega_0)$.

Upon tracing out the atom, we obtain for the signal mode $s$ the ME 
\begin{eqnarray}  \label{pis-me}
\dot{\rho}_s&=& g_h^2 \Big[|I_{h,gi}|^2 (\bar{n}_h+1) \rho_{ee}([b\rho_s,
b^\dagger]+[b,\rho_s b^\dagger])  \notag \\
&+& |I_{h,ei}|^2 \bar{n}_h\rho_{gg}([b^\dagger\rho_s, b]+[b^\dagger,\rho_s
b]) \Big],
\end{eqnarray}
where we have assumed for simplicity that $|I_{h,gi}| = |I_{h,ei}|.$

The resulting time evolution of the signal-mode Fock state $n_s$ is given by 
\begin{equation}
\dot{n}_s = - 2 g_h^2 |I_{h,gi}|^2 \left((\bar{n}_h+1) n_s \rho_{ee} - \bar{n%
}_h (n_s+1) \rho_{gg} \right),
\end{equation}
%the work stored  in the following form \cite{Ghosh2018}.
For the Glauber-Sudarshan P-distribution of the signal state, i.e., 
%\begin{equation}
$\rho_s = \int P(\alpha) |\alpha\rangle \langle \alpha | d^2 \alpha$, 
%\end{equation}
one obtains the Fokker-Planck (FP) equation 
\begin{equation}
\frac{\partial}{\partial t} P(\alpha) = -\frac{\mathcal{G} }{2} \left( \frac{%
\partial}{\partial \alpha} + \frac{\partial}{\partial \alpha^\star} \right)
P + \mathcal{D} \frac{\partial^2 P}{ \partial \alpha \partial \alpha^\star},
\end{equation}
with 
\begin{eqnarray}
\mathcal{G} & = & \frac{2 g_h^2 |I_{h,ai}|^2 (n_h-n_c)}{2n_c+1}  \notag \\
\mathcal{D} & = & \frac{2 g_h^2 |I_{h,ai}|^2 n_h (n_c +1)}{2n_c + 1}.
\end{eqnarray}
Here $\mathcal{G}$ describes the effective gain rate in the amplification
regime and $\mathcal{D}$ describes the diffusion rate for the process. An
initial coherent state $|\alpha_0\rangle$ then evolves into \rr{displaced thermal state}
\begin{equation}
P(\alpha, t) = \frac{1}{\pi \sigma^2 (t)} Exp \left( -\frac{|\alpha -
\alpha_0 e^{\mathcal{G}t/2}|^2}{\sigma^2 (t)} \right),
\end{equation}
with $\sigma^2 (t) = \mathcal{G}/\mathcal{D} (e^{\mathcal{G}t} -1)$ \rr{and $\alpha_0 e^{\mathcal{G}t/2}$ being the displacement from the phase-space origin. The energy of the signal mode $\rho_s$ reads as
\begin{equation}
    E(\rho_s(t))= \nu[|\alpha_0|^2 e^{\mathcal{G}t}+\sigma^2 (t)],
\end{equation}
where the first term corresponds to the energy associated with the displacement and the second term corresponds to the thermal energy. By a displacement operator, one can bring the state to the phase-space origin and thereby extract $\nu[|\alpha_0|^2 e^{\mathcal{G}t}$ amount work \cite{WOF}. Therefore, the rate of change of ergotropy of the evolved state is given by 
\begin{equation}
\dot {\mathcal{W}}  = \nu[|\alpha_0|^2 \mathcal{G}e^{\mathcal{G}t},
\end{equation}
which quantifies the power of the coherent amplification.}

\rr{The efficiency then evaluates to 
\begin{equation}
 \eta = \frac{\nu }{\Omega _{h}}\frac{|\alpha _{0}|^{2}}{|\alpha_0 |^{2}+\frac{n_{h}(n_{c}+1)}{n_{h}-n_{c}}}.
\end{equation}
For $n_c =0$ ($T_c=0$) we retrieve the result in the main text.}

\section{\label{AppendixE} The three-body interaction and Raman Hamiltonian}
\begin{figure}[th]
\centering
\includegraphics[scale=0.5]{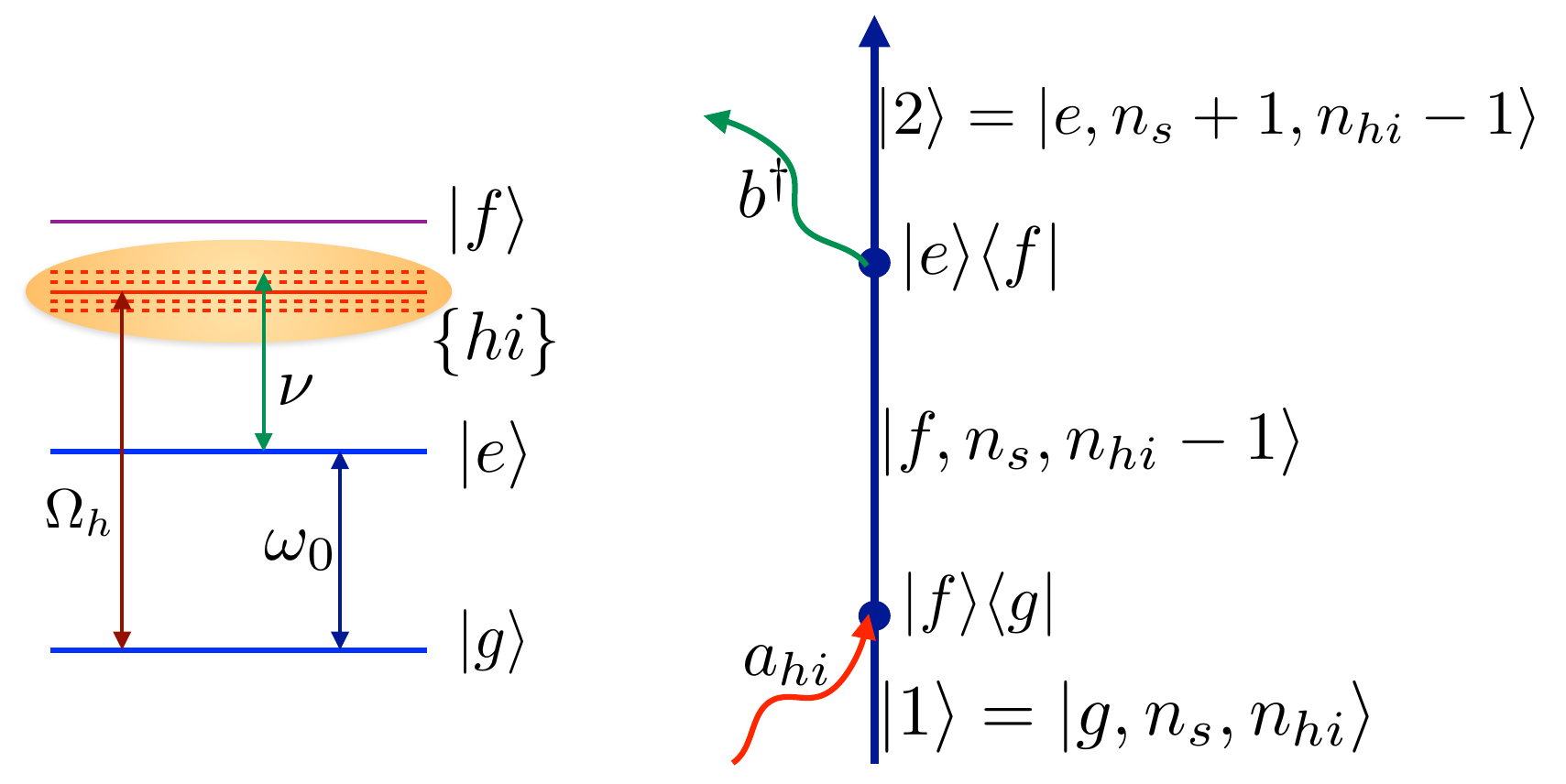}
\caption{Schematic of the three-body interaction via a two-photon Raman process.}
\label{Raman}
\end{figure}
\rr{ One possible way to visualize the three-body interaction in Eq. (\ref{sc1}), by two two-body transitions is via a two-photon Raman transition (see Fig. \ref{Raman} ). The transition from $|1\rangle=|g,n_{s},n_{hi}\rangle \rightarrow |2\rangle=|e,n_{s}+1,n_{hi}-1\rangle$ can be considered as a two-step process as shown in Fig. \ref{Raman}. The Raman Hamiltonian $H_R$ should result in a same transition by a one-step process. The effective Raman Hamiltonian should then be given by
\begin{equation}
    H_R= \sum_{i}g_{hi}\left(\phi_{hi} \hat{a}_{hi}\hat{b}^{\dagger
}|e\rangle \langle g|e^{-i[\Omega _{hi}-(\nu +\omega _{0})]t}+H.c.\right),
\end{equation}
in the interaction picture. The coupling strength is given by \cite{Ghosh2018} 
\begin{equation}
  g_{hi}= 2\pi \frac{g_{ef}g_{fg}}{E_f-E_g-\Omega_{hi}},  
\end{equation}
$g_{ij}$ being the dipolar coupling constant of the atom between the levels $i$ and $j$.}
\newpage
%\bibliography{BH}

\providecommand{\noopsort}[1]{}\providecommand{\singleletter}[1]{#1}%

\end{document}